\documentclass[pre,aps,twocolumn]{revtex4}

\usepackage[normalem]{ulem}

\usepackage[english]{babel}
\usepackage{graphicx}
\usepackage{amssymb}
\usepackage{amstext}

\usepackage{url}

\begin{document}

\title{Lower glycolysis carries a higher flux than any biochemically possible alternative}


\author{Steven J. Court, Bart\l omiej Waclaw, Rosalind J. Allen}
\affiliation{SUPA, School of Physics and Astronomy, University of Edinburgh, Mayfield Road, Edinburgh EH9 3JZ, United Kingdom }



\begin{abstract} 
The universality of many pathways of core metabolism suggests a strong role for evolutionary selection, but 
it remains unclear whether existing pathways have been selected from a large or small set of biochemical possibilities. 
To address this question, we construct {\em{in silico}} all possible biochemically feasible alternatives to the   trunk pathway of glycolysis and 
gluconeogenesis, one of the most highly conserved pathways in metabolism. 
We show that, even though a large number of alternative pathways exist, the alternatives carry lower flux than the real pathway under typical 
physiological conditions. Alternative pathways that could potentially carry higher flux often lead to infeasible intermediate metabolite concentrations.
We also find that if physiological conditions were different, different 
pathways could outperform those found in nature. Our results demonstrate how the rules of biochemistry restrict the alternatives that are 
open to evolution, and suggest that the existing trunk pathway of glycolysis and gluconeogenesis represents a maximal flux solution. 
\end{abstract}

\maketitle

\section{Introduction}

The biochemical pathways of central carbon metabolism are highly conserved across all domains of life, and largely control the productivity 
of life on Earth \cite{Romano1996,EricSmith2004}. Yet it remains unknown whether these pathways are the result 
of historical contingency during early evolution \cite{Pal2006}, or  are instead optimal solutions to the problem of energy and biomass 
production \cite{PentoseGame, Ebenhoeh2001, Heinrich97, Flamholz2013,Wagner2013,Noor2010}. Put simply, are there alternative 
biochemically feasible pathways that could perform the same function, and if so, how do they perform compared to those found in nature?  

Previous studies have mainly addressed this question either by constructing simplified artificial metabolic networks \cite{PentoseGame, Ebenhoeh2001, Heinrich97}, 
or by mining databases of  biochemical compounds and reactions for known organisms \cite{Flamholz2013,Wagner2013,Samal2011,Bar-EvenPNAS,HandorfExpanding}. Both of these approaches 
have drawbacks: the former does not capture real biochemistry, while the latter is limited to metabolites and reactions found in well-studied organisms.  Nevertheless this work 
has led to suggested optimality principles including maximizing biochemical flux or yield \cite{Ebenhoeh2001, Heinrich97},   minimizing biochemical steps or protein 
costs \cite{PentoseGame,Flamholz2013}, and ensuring that function is maintained  in a changing 
environment \cite{Samal2011}. Only a few studies have attempted to explore the full universe of possible  metabolic pathways, using realistic rules of 
 biochemistry \cite{Melendez-Hevia97,Mittenthal1998,Hatzimanikatis2005,Henry2010,Noor2010}. 
 In particular, using this approach Noor et al. \cite{Noor2010} recently suggested that central carbon metabolism, taken as a whole, 
uses the minimal number of enzymatic steps needed to generate a predefined set of  biochemical precursors.

\begin{figure}[!b]
  \centering
  \includegraphics[width=0.45\textwidth]{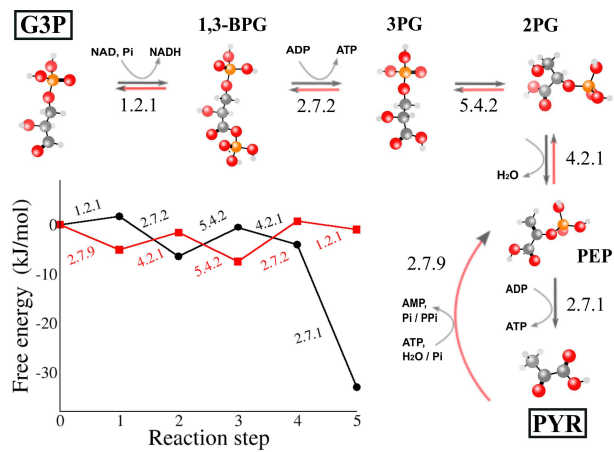}
  \caption{The trunk pathway of glycolysis and gluconeogenesis. 
The end points are glyceraldehyde 3-phosphate (G3P) and pyruvate (PYR); intermediate metabolites are  1,3-bisphosphoglycerate (1,3-BPG); 
3-phosphoglycerate (3PG); 2-phosphoglycerate (2PG); 
and phosphoenolpyruvate (PEP). For each reaction the external metabolites  involved and the first 3 numbers from the EC number classification are 
indicated.  
The inset shows thermodynamic profiles for the trunk pathway in the glycolytic and gluconeogenenic (pps) directions (see Methods).
\label{fig:TrunkPathway}}
\end{figure}

Here, we perform an exhaustive computational search for all possible biochemically feasible  alternatives to one of the most ancient and most highly conserved 
sections of central carbon metabolism, the trunk pathway of glycolysis and gluconeogenesis. Glycolysis breaks down glucose to pyruvate, generating ATP, NADH and 
biosynthetic precursors, while gluconeogenesis 
 uses ATP and NADH to generate glucose from  pyruvate. 
The glycolytic-gluconeogenic pathway (Fig.~S1) is almost linear and can be divided into  two parts: an ``upper'' chain of reactions 
involving 6-carbon molecules, which connects glucose to glyceraldehyde-3-phosphate (G3P), and a ``lower'' chain of reactions involving 3-carbon molecules, 
 which connects G3P to pyruvate  (Fig.~\ref{fig:TrunkPathway}). This lower reaction chain is known as the trunk pathway. 
In prokaryotes,  the glycolytic and gluconeogenic trunk pathways consist of almost the same set of 5 reactions, differing only in one step. In  glycolysis, the exergonic conversion of 
phosphoenolpyruvate (PEP) to pyruvate  is coupled to the phosphorylation of ADP to ATP, while in gluconeogenesis  
 the reverse reaction is driven by the hydrolysis of ATP  to AMP, either with release of  inorganic phosphate (the phosphoenolpyruvate 
synthase (pps) route), or via consumption of inorganic phosphate and release of pyrophosphate (the phosphate dikinase (ppdk) route).

While the upper part of the glycolytic/gluconeogenic pathway exists in several very distinct variants, most notably the Embden-Meyerhof-Parnas 
and Entner-Doudoroff pathways (Fig.~S1) \cite{Flamholz2013, Bar-Even2012}, the trunk pathway is ubiquitous, 
and contains enzymes which are highly conserved and universally distributed across the three domains of life \cite{Ronimus2002, Verhees2003}. 
The existence of such an ancient and universal pathway suggests three possible scenarios: (i) the trunk pathway is the only biochemical possibility, (ii)  
alternatives exist but the extant pathway is evolutionarily optimal and (iii) alternatives are possible but have not been found by evolution. Distinguishing 
these scenarios lies at the heart of much of evolutionary biology. Here, we address this question directly using a computational approach. 
By systematically constructing and exploring the full space of biochemically feasible metabolites and reactions, many of which are not 
currently exploited by any known organism, we find that hundreds of alternative trunk pathways are possible. The one observed in 
nature, however, carries the maximal biochemical flux, under reasonable constraints on the intermediate metabolite and enzyme concentrations. 
Our results suggest that the trunk pathway represents an optimal solution among many possible alternatives.

\section{Results}

\subsection{A network of all possible biochemical reactions}
We used a  computer program to generate the network of  all possible biochemically feasible pathways between G3P and pyruvate. 

{\em{Metabolites.}} Our program first systematically generates all possible relevant {\em{internal metabolites}}, i.e. molecules intermediate between G3P and pyruvate,
including those that are not found in nature. 
Central carbon metabolism consists exclusively of reactions between ``CHOP'' molecules: those composed of carbon, hydrogen, oxygen and phosphorus atoms, with the latter 
being present only in phosphate groups. Nitrogen and sulfur are present only in cofactors such as ATP, NAD and CoA \cite{MorowitzPNAS, MorowitzCOMPLEXITY}. 
Moreover, the trunk pathway contains only unbranched aliphatic 3-carbon CHOP molecules, with the exception of the 4-carbon oxaloacetate, used in
gluconeogenesis in liver and kidney cells \cite{Stryer}.
We therefore include in our analysis all possible unbranched aliphatic CHOP molecules 
containing up to 4 carbon atoms. We consider only molecules which are electrostatically charged (i.e. include carboxyl or phosphate groups). This condition 
is motivated by the need to avoid leakage through the lipid membrane, and is satisfied by almost all molecules in core metabolism \cite{Srinivasan2009}. 
Applying these criteria results in over 1000 different molecules, including all the internal metabolites in the real trunk pathway.
 We computed the free energy of formation $\Delta_f G $ for all our internal metabolites, using
existing experimental data where possible \cite{alberty_biochemical_2006} or, in the absence of such data, using a variant of the group contribution 
method \cite{Mavrovouniotis91, Mavrovouniotis90, Jankowski2008} (see Methods and Supporting Information \cite{suppmatt}). 

{\em{Reactions.}} Our program generates all possible reactions among our set of internal metabolites, based on 12 EC reaction classes \cite{EC}, which 
 encompass all the reactions between  CHOP molecules of length 2-4 carbons in core metabolism 
(Table \ref{tb:reactions}) \cite{Stryer}. Many of these reactions involve cofactors such as  ATP and NAD; we refer 
to these as \textit{external metabolites} and we assume that they have fixed concentrations, which define the cellular (physiological) conditions. 
The external metabolites in our reaction network are ATP, ADP, AMP, NAD, NADH, inorganic phosphate Pi, pyrophosphate PPi, CO$_2$ and H$_2$O. 
For each of our reaction classes, we include all  known couplings  with the external metabolites (see Supporting Information \cite{suppmatt}). 

\begin{center}
\begin{table*}[!t]
{\small
\hfill{}
\begin{tabular}{cll}
\hline
\textbf{EC class}&&\textbf{Example reaction}\\
\hline
1.1.1              & oxidation                              & CH$_3$-CH$_2$(OH) + NAD$^+$ $\rightleftharpoons$ CH$_3$-CHO + NADH \\
1.2.1              &             & CH$_3$-CHO + NAD$^+$ + Pi $\rightleftharpoons$ CH$_3$-COp + NADH + H$^{+}$ \\ 
\hline
2.7.1              &                                & CHO-CH$_2$(OH) + ATP $\rightleftharpoons$ CHO-CH$_2$p + ADP \\
2.7.2              & phosphate transfer                  & CH$_3$-COOH + ATP $\rightleftharpoons$ CH$_3$-COp + ADP \\
2.7.9              &                                   & CH$_3$-CO-COOH + ATP + H$_2$O/Pi $\rightleftharpoons$ CH$_2$=Cp-COOH + AMP + Pi/PPi \\
\hline
3.1.3              & hydrolysis     & CH$_3$-CH$_2$p + H$_2$O  $\rightleftharpoons$  CH$_3$-CH$_2$(OH) + Pi  \\
\hline
4.1.1              & decarboxylation                      & CH$_3$-CH(OH)-COOH + H$_2$O $\rightleftharpoons$ CH$_3$-CH$_2$(OH) + CO$_2(aq)$  \\
\hline
4.2.1              & dehydration                         & CH$_3$-CH(OH)-COOH $\rightleftharpoons$ CH$_2$=CH-COOH + H$_2$O\\
\hline
5.3.1              & isomerisation        & CH$_3$-CO-CH$_2$(OH) $\rightleftharpoons$ CH$_3$-CH(OH)-CHO \\
\hline
5.3.2              & tautomerism            & CH3-C(OH)=C(OH)-CH$_2$p $\rightleftharpoons$ CH$_3$-CO-CH(OH)-CH$_2$p\\
\hline
5.4.2              & isomerisation                 & CH$_3$-CH(OH)-CH$_2$p $\rightleftharpoons$ CH$_3$-CHp-CH$_2$(OH)\\
\hline
6.4.1              & ATP-driven carboxylation                     & CH$_3$-CH$_2$(OH) + CO$_2(aq)$ + ATP $\rightleftharpoons$ CH$_3$-CH(OH)-COOH + ADP + Pi\\
\hline
\end{tabular}}
\hfill{}
\caption{The set of reaction types included in our analysis, defined by the first 3 numbers of the EC classification. 
Phosphate groups are denoted by a ``p''. Not all variants of each reaction type are listed here; for a full list see Table SIII. \label{tb:reactions}}
\end{table*}
\end{center}

\subsection{Our network generates many alternative trunk pathways}
Our network reveals a huge number of alternative pathways connecting  G3P and pyruvate, which 
 are consistent with the rules of biochemistry. For example, in the glycolytic direction, we find 13 pathways of length 4, 
532 pathways of length 5, and the number of pathways increases 
exponentially with the path length (Fig.~S2). 
Some of these alternative pathways use the same set of
reaction types as  the real trunk pathway, but execute them in a different order
(e.g. Fig.\ref{fig:Fig3}~E, left). Others make use of a different set of reaction types (Fig~\ref{fig:Fig3}E, right).
 A similar picture holds in the gluconeogenic direction (Fig.~S2).

How many of these alternative trunk pathways are feasible under typical physiological conditions? We first demand that candidate pathways in 
the glycolytic direction should produce at least 2 ATP molecules. This is required to make the whole glycolytic pathway produce a net ATP yield, since the two dominant forms of 
glycolysis, the EMP and ED pathways, consume 1 ATP molecule per G3P  in their upper halves.  In the gluconeogenic direction, the real trunk pathway consumes 2 ATPs, 
one of which is converted to ADP and the other to AMP (Fig.~\ref{fig:TrunkPathway}); the AMP being eventually recycled to ADP via the consumption of a third 
ATP in the adenylate kinase reaction. 
 In this direction, we exclude any candidate paths that have a poorer  yield of G3P per ATP than the real path. 
Motivated  by considerations of simplicity \cite{PentoseGame, Noor2010} 
and cost of enzyme production \cite{Flamholz2013}, we then further restrict our analysis to those pathways with the minimum number of steps. In both the glycolytic and 
gluconeogenic directions, the minimum pathway length consistent with the above requirements is 5 steps;
our network generates 202 candidate 5-step glycolytic paths, and 300 candidate 5-step gluconeogenic paths.

\subsection{Real glycolysis and gluconeogenesis carry maximal flux under physiological conditions}
We next evaluated the performance of the alternative trunk pathways generated by our network, by comparing their steady-state metabolic flux.
For glycolytic pathways, the metabolic flux corresponds to the  rate 
of ATP production, while for gluconeogenic pathways, it corresponds to  the rate of production of G3P (and ultimately of glucose). This flux  depends not only 
on the total free energy change across a given pathway, but also on the distribution of 
the individual reaction free energies (see e.g.~Fig.~\ref{fig:TrunkPathway}), which in turn  depends on the  intracellular 
environment via the
concentrations of the external metabolites. For linear pathways, the flux can be calculated analytically, assuming linear kinetics with 
diffusion-limited enzymes \cite{Heinrich97, Albery76, Heinrich91} (see Methods). We also assume that, for each pathway, the individual enzyme 
concentrations are optimised to maximise pathway flux, for a fixed total amount of enzyme (although similar results are obtained when we relax this 
assumption; Fig.~S3).

\begin{figure*}[!t]
  \centering
  \includegraphics[width=\textwidth]{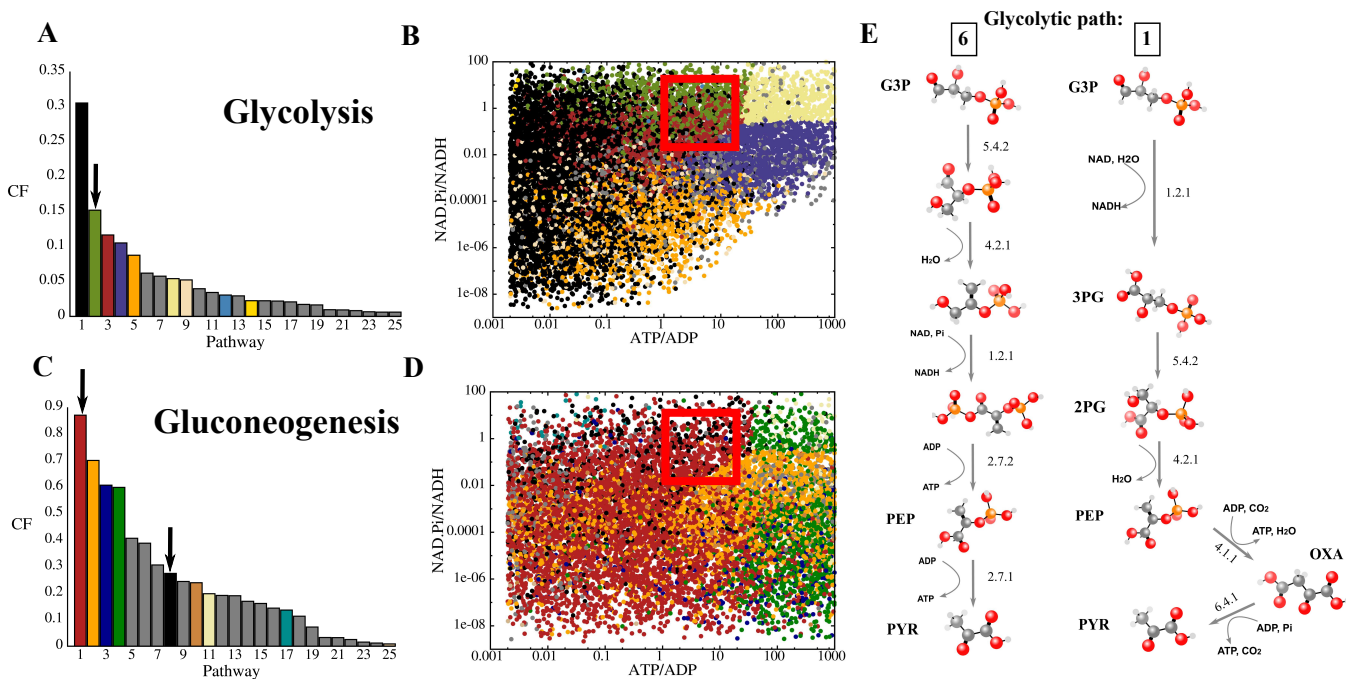}
  \caption{The real glycolytic and gluconeogenic trunk pathways represent maximal flux solutions. 
A, C: Alternative pathways generated in our analysis ordered by their comparative flux (see text), averaged across the whole 10-dimensional parameter space. 
The top 25 paths are shown. 
Panel A shows glycolytic pathways; the green bar (indicated by the arrow) is the real pathway. Panel C shows gluconeogenic paths; the black and red 
bars (indicated by arrows) are the real pps and ppdk routes.
B, D: Relative pathway performance as a function of the intracellular environment. Each dot represents a randomly sampled point in parameter space; the colour of the dot 
indicates the candidate pathway which had the highest flux at that point in parameter space (colours as in panels A,C). For parameter sampling 
procedure, see Methods. The axes represent the redox state and energy state of the cell: [NAD][Pi]/[NADH] and [ATP]/[ADP]. 
The red boxes indicate the typical 
physiological state of the cell \cite{Berg2009,Bennett2009}: [ATP]/[ADP]  = 3-20, [NAD][Pi]/[NADH] = 0.01-10M  [based on [NAD]/[NADH] =10-100 and [Pi] = 1-100mM].
Panel B shows glycolytic 
pathways; real glycolysis (green) does best under typical physiological conditions. Panel D shows gluconeogenic pathways; the two real routes (red and black) perform 
best under typical cellular conditions. 
E: two alternative glycolytic pathways, numbers 6 and 1.  
In the molecular representations, grey, white, red and orange spheres represent carbon, hydrogen, oxygen and phosphorus atoms respectively. 
\label{fig:Fig3}}
\end{figure*}

Importantly, when calculating the metabolic flux, we impose constraints on the intermediate internal metabolite concentrations. For the 
bacterium \emph{E.coli}, metabolite concentrations 
range from 0.1$\mu$M to 100mM, with the total intracellular metabolite pool being around 300mM \cite{Bennett2009}. We expect that  very low 
metabolite concentrations are undesirable due to molecular noise, while very high concentrations are precluded by osmotic considerations. We therefore 
set the flux to zero for a given pathway if any of its intermediate metabolite concentrations falls outside the range  1nM to 0.5M.  

Imposing this constraint, we calculated the metabolic flux of all our  202 glycolytic, and 
300 gluconeogenic candidate paths, across a wide range of intracellular conditions, as defined by the external metabolite concentrations. We randomly selected 
10,000 points from the 10-dimensional parameter space consisting of the 
concentrations of the 8 external metabolites, G3P and pyruvate, sampling each parameter logarithmically over several orders of magnitude above and below its 
typical physiological concentration (see Supporting Information \cite{suppmatt}). For each point in parameter space we evaluated the flux of each candidate pathway.

As a simple metric, we first compare the performance of our candidate pathways, averaged over the entire parameter space. To this end, 
we compute the \textit{comparative flux} (CF), which is the flux of a given pathway, divided by the  maximum flux 
obtained amongst all pathways, at a given point in parameter space. Averaging this quantity across the whole parameter 
space gives a measure of relative performance, for each candidate path. We find that the real glycolytic and gluconeogenic 
pathways perform remarkably well, compared to the many alternatives (Fig.~\ref{fig:Fig3}, left panels). For glycolysis (Fig.~\ref{fig:Fig3}A), 
the natural trunk pathway (in green, indicated by the arrow) outperforms all the alternative pathways except one. For gluconeogenesis, 
the two natural variants, with different cofactor coupling for the pyruvate to PEP step, are ranked first (the pps route, shown in 
red in  Fig.~\ref{fig:Fig3}C) and 8th (the ppdk route, shown in black in  Fig.~\ref{fig:Fig3}C). These results strongly suggest that 
the natural trunk pathways carry a high flux compared to alternatives; however this metric is dependent on the range of parameter space that is sampled.

To investigate  in more detail, we analyzed which of our candidate pathways achieved the highest flux at different points in the parameter space. This allows us to understand how the  performance of a given pathway depends on the intracellular environment. Our results show that different candidate pathways perform best in different regions of the parameter space. In particular, pathway performance is very sensitive to the cellular energy state, as measured by  the ratio of the ATP and ADP concentrations ([ATP]/[ADP]), and redox state, as measured by the ratio [NAD][Pi]/[NADH] (Fig.~\ref{fig:Fig3} (middle panels)). Focusing on the glycolytic pathways (Fig.~~\ref{fig:Fig3}B) we see that, remarkably, the natural trunk pathway (green dots) 
outperforms all the alternatives in the region of parameter space close to that found in living cells (red box). This suggests that the glycolytic trunk pathway represents a 
maximal flux solution for the conversion of G3P to pyruvate, under typical intracellular conditions. For the gluconeogenic pathways (Fig.~\ref{fig:Fig3}D), a similar picture holds. Here, the two pathways found in nature, the pps-route (red) and the ppdk-route (black), both outperform the alternatives under typical physiological conditions (red box). The relative performance 
of these two pathways depends sensitively on the concentration of pyrophosphate (see Fig.~S4).

\subsection{Alternative trunk pathways} 
\begin{figure}[!t]
  \centering
  \includegraphics[width=0.45\textwidth]{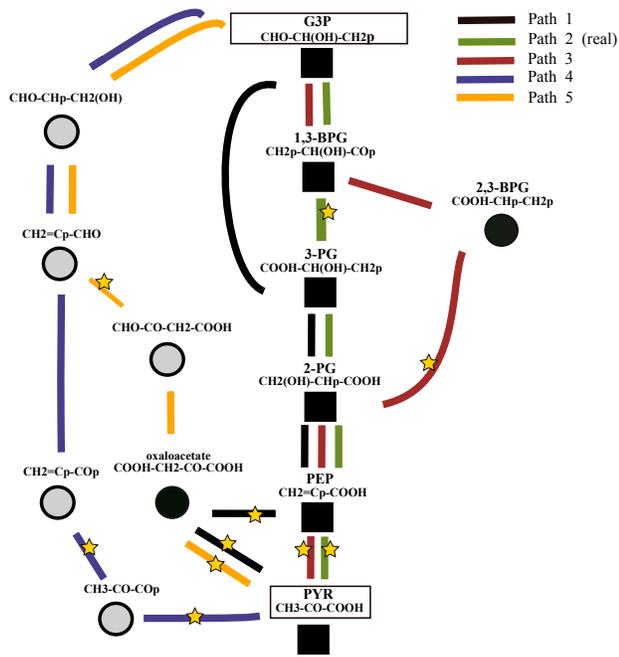}
\caption{Alternative pathways which perform well in the glycolytic direction. Reactions are indicated by lines; the colour denotes the 
pathway as in Fig.~\ref{fig:Fig3}A and B; stars indicate ATP-producing steps. 
Intermediate metabolites are shown by symbols; black squares indicate 
metabolites in the real trunk pathway, black and grey circles indicate metabolites present and not present in the KEGG database  \cite{KEGG}, respectively.
\label{fig:Glyc_schematic}}
\end{figure}

Our analysis reveals several alternative pathways that can, under different intracellular conditions, 
outperform the true glycolytic and gluconeogenic pathways (Fig.~\ref{fig:Fig3}, middle panels and Fig.~\ref{fig:Glyc_schematic}, Fig.~S5-S8). 

In the glycolytic direction, path 1 (black) outperforms the real trunk pathway for low [ATP]/[ADP] ratios (Fig.~\ref{fig:Fig3}B), resulting in its 
apparently better performance than the real pathway when averaged over the whole parameter space (Fig.~\ref{fig:Fig3}A). This pathway, shown 
on the right of Fig.~\ref{fig:Fig3}E, differs from the real pathway in that it converts G3P directly to 3-phosphoglycerate (3-PG) without the production of 
ATP (Fig.~\ref{fig:Glyc_schematic}, compare black and green). This means that it has a  highly exergonic oxidation reaction as its first 
step. In a linear pathway, the  initial reactions tend to exert the greatest control over the flux \cite{Heinrich97}, so an exergonic first reaction 
can result in a large flux. However, pathway 1 is highly sensitive to the constraints on the intermediate metabolite concentrations. Because of 
its exergonic first reaction, it tends to accumulate high concentrations of downstream metabolites, causing it to be deemed infeasible in 
our analysis over a large part of the parameter space.

Pathway 3 (red), 
is similar to the real glycolytic trunk path, except that 1,3 bisphosphoglycerate (1,3-BPG) is first isomerized (to 2,3-BPG) and then dephosphorylated 
(to 2-phosphoglycerate 2PG, with ATP generation), rather than being first dephosphorylated and then isomerized as in the real pathway (Fig.~\ref{fig:Glyc_schematic} and Fig.~S5). 
The large free energy change for the 
isomerization ($\Delta G=-28$kJmol$^{-1}$) means that pathway 3 can in principle carry a higher flux than the real pathway, 
but because of this large drop in free energy it tends to accumulate high concentrations of 2,3-BPG 
 and is therefore only favourable for low concentrations of the starting metabolite G3P. 
Interestingly, a 
similar pathway exists in red blood cells, where 2,3-BPG is produced from 1,3-BPG via the Rapoport-Leubering shunt \cite{Cho23bpg}. In red blood cells, however, 
the 2,3-BPG is hydrolysed to either 3-PG or 2-PG without ATP generation \cite{Cho23bpg}, thus sacrificing one ATP compared to the usual glycolytic pathway. 
It is tempting to hypothesize that nature is forced to sacrifice an ATP molecule when using this shunt, to prevent the buildup of 2,3-BPG, 
which is already the most concentrated organophosphate in erythrocytes. 

Pathway 4 (violet) outperforms the real pathway at high  ATP concentrations (Fig.~\ref{fig:Fig3}B); this is because its ATP producing steps are at the 
end (4th and 5th steps), in contrast to the real pathway where ATP is produced in the 2nd and 5th steps (Fig.~\ref{fig:Glyc_schematic}, Fig.~S5). Because later steps in a linear 
pathway tend to have less impact on the flux \cite{Heinrich97}, this makes pathway 4 more tolerant of high  ATP concentrations than the real pathway. Similarly, 
pathway 5 (orange) differs from the real pathway in that its oxidation step is moved to the end (the 4th step rather than the first as in the real pathway); 
this makes pathway 5 more tolerant of reducing conditions than the real pathway (Fig.~\ref{fig:Fig3}B, Fig.~S5). Several other interesting 
alternative glycolytic trunk pathways are discussed in the Supporting Information \cite{suppmatt}, section XI. 

In the gluconeogenic direction (Fig.~\ref{fig:Fig3}, C and D; Fig.~S7 and S8), pathways 1 (red) and 8 (black) correspond to the two prokaryotic trunk 
pathway variants found in nature, the pps and ppdk routes. Interestingly, alternative pathways 2-7 
all contain the same set of reaction types as the real pps pathway, but carry out these reactions in varying order, thus making use of different intermediate internal 
metabolites. This affects their relative performance in different regions of the parameter space. For example, pathway 2 (orange) and pathway 4 (green) both differ 
from the real pps pathway in that ATP is consumed in the first two reactions (rather than in the 1st 
and 4th reactions as in the real pps pathway, Fig.~S8). This makes their flux more sensitive to the ATP concentration, explaining why they dominate at high ATP/ADP 
ratio (Fig.~\ref{fig:Fig3}D). 

Interestingly, eukaryotes use a slightly different gluconeogenic trunk pathway, in which the conversion of pyruvate to PEP is a two-step 
process (via 4-carbon oxaloacetate) in which ATP is converted to ADP twice, rather than a 1-step process, converting ATP to AMP as in the prokaryotic 
pps and ppdk routes. Because the eukaryotic gluconeogenic trunk pathway is of length 6 steps, it was not considered in our analysis. However, 
when we repeat our analysis in the absence of dikinase reactions ({\em{i.e.}} not allowing the conversion of ATP to AMP), we find that 
the shortest feasible gluconeogenic pathways are 6 steps, and that
the real eukaryotic pathway outperforms all alternatives  under physiological conditions (Fig.~S9).

\subsection{Constraints on metabolite concentrations are important} 
\begin{figure}[t]
  \centering
  \includegraphics[width=0.49\textwidth]{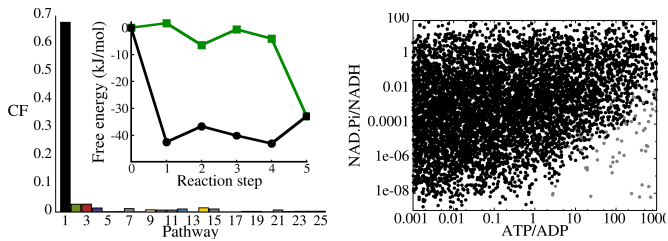}
\caption{Constraints on metabolite concentrations are crucial. Repeating our analysis without these constraints, in the glycolytic 
direction, pathway 1 (black in Fig.~\ref{fig:Fig3}A and B and shown on the right of Fig.~\ref{fig:Fig3}E) dominates over the whole parameter space. 
The inset to the left panel shows the thermodynamic profile for pathway 1 (black) alongside that of the real path 2 (green). \label{fig:Glyc_noConcRest} }
\end{figure}

Imposing constraints on the intermediate internal metabolite concentrations is crucial to 
our analysis. Repeating our analysis without these constraints produces a very different outcome (Fig.~\ref{fig:Glyc_noConcRest}).
 In the glycolytic direction,  
if the metabolite concentrations are unconstrained, glycolytic path 1 (Fig.~\ref{fig:Fig3}E, right; black in Fig.~\ref{fig:Fig3}A and B), which has a highly 
exergonic first reaction, produces the highest flux across the entirety of 
the parameter space. A similar picture emerges in the gluconeogenic direction (Fig.~S10).
The fact that our results are strongly affected by constraining the metabolite concentrations  highlights the importance of considering metabolite 
concentrations when using methods such as flux balance analysis \cite{Varma1994} to study metabolic networks. 

Interestingly, all the enzymes required for glycolytic pathway 1 exist in nature, in various organisms (Fig.~S6), although this route 
is not known to be used in nature as a glycolytic  pathway. Our analysis hints that constructing this pathway could provide an interesting 
target for synthetic biology, since it could, in principle, provide a new way to accelerate growth of organisms useful for biotechnological 
applications, if suitable branching pathways were provided to prevent the buildup of downstream metabolites.

We also repeated  our analysis, imposing a {\em{narrower}} concentration range for the intermediate metabolite concentrations ($10^{-7}$ to $10^{-2}$M 
rather than $10^{-9}$ to $5 \times 10^{-1}$M). Under these constraints, we find that the real glycolytic pathway outperforms  the alternatives over a wider 
range of parameter space, while the restricted concentration range has little effect on our results for the gluconeogenic paths (Fig.~S11). This 
further supports the picture emerging from our analysis, that in the presence of reasonable limits on intermediate metabolite concentrations, the real glycolytic 
and gluconeogenic paths outperform the alternatives.

\section{Discussion}

Despite the huge variety and complexity of life on Earth, the biochemistry of core metabolism is remarkably universal. Our analysis 
shows that this universality does not arise from an absence of other possibilities. Using a systematic approach, we have identified many alternatives to perhaps 
the most highly conserved set of metabolic reactions, the glycolytic and gluconeogenic trunk pathways. Our alternative pathways obey 
the rules of biochemistry,  carry positive flux under reasonable intracellular conditions, and satisfy reasonable constraints on metabolite 
concentrations. Remarkably, of all these alternatives, we find that the  trunk pathway observed in nature carries the highest biochemical flux 
in both the glycolytic and gluconeogenic directions, for parameters that represent typical 
intracellular physiological conditions. Of the two variants of the prokaryotic gluconeogenic pathway that are found in nature, the pps route is 
the best performer across a wide parameter range, while the ppdk route also carries a high flux, but is more sensitive to environmental conditions, 
requiring a low concentration of pyrophosphate (Fig.~S4). The fact that our analysis identifies the real pathways using only flux maximization combined with 
constraints on intermediate  metabolite concentrations and a requirement for minimal pathway length suggests that these three factors are all likely 
to have been important driving forces in the evolution of metabolism. 

Flux maximization is widely recognized as an important concept in the study of metabolism; both from the perspective of glycolysis 
\cite{Heinrich97, Ebenhoeh2001} and more broadly  \cite{Varma1994}. Our results support this picture, but  suggest that evolutionary pressures 
on metabolic fluxes have to operate within the context of reasonable constraints on metabolite concentrations, and that neglecting these constraints can produce 
dramatically different outcomes. Our results 
expand on  the recent suggestion of Noor et al.~\cite{Noor2010} that central carbon metabolism can be understood as a minimal walk between 
the set of metabolites essential for growth. In our analysis, the real pathways do indeed minimize the total number of reaction steps; this 
imposes a strong constraint on the number of alternative paths. However we find that the requirement to produce a set of essential biochemical 
precursors is not sufficient to explain the biochemical structure of the natural trunk pathway. 
Firstly, alternative pathways are possible which produce the essential precursors with the same number of steps (e.g.~glycolytic path 1). 
Secondly, many of our alternative pathways produce 
very similar, but not identical, intermediates to those of the real trunk pathway and it is conceivable that these could be used as alternative precursors.
Our results show that flux maximization provides a criterion by which these alternative minimal-length pathways may be distinguished.  

Our analysis also reveals alternative trunk pathways which can perform better than the real one under different concentrations of reactant, product 
and external metabolites. While some of these alternatives involve compounds and reactions which are not found in biochemical databases, others use 
enzymes which are known to exist in nature. 
The latter pathways are clearly plausible biochemically but are apparently not used in nature. In some cases (e.g. our alternative 
glycolytic paths 1 and 3) this is probably because they tend to generate large intermediate concentrations. In other cases (e.g. glycolytic
pathways 4 and 5), the alternatives are not optimal under typical physiological conditions, but would be optimal if conditions were different. 

In this study, we have limited our analysis to pathways which start and end at G3P
and pyruvate. Relaxing this requirement would certainly lead to many more alternative pathways for the  generation of energy, 
and for biosynthesis. 
While it is also important to consider other factors, including the need for integration within a wider metabolic network,
our analysis suggests that key principles underlying the structure of core metabolism may emerge from 
simple biochemical, thermodynamic and biophysical considerations.

\section{Methods}

\subsection{Chemical compounds and reactions}
We created a list of chemical compounds with 2, 3 or 4 carbon atoms by generating all possible linear combinations of the 17 ``building blocks''
shown in Table SI. 
Each of the building blocks was composed of a single carbon atom with associated oxygen, hydroxyl, hydrogen and/or phosphate groups. Building blocks were 
connected together in linear chains by single or double bonds.
This procedure created 1008 linear molecules, 828 of which are electrostatically charged in solution,  i.e.~containing at least one carboxyl or phosphate group. 
These 828 molecules are our internal metabolites. 
Next, for every possible pair of molecules from this list we checked systematically whether the reactions from Table \ref{tb:reactions} could transform one molecule 
into another, allowing for all possible couplings with the external metabolites. In this way, a network of 7145 reactions was generated (see Supporting Information \cite{suppmatt} section VII).

\subsection{Free energies of compounds and reactions}
\,For those internal metabolites which are known biochemical species, standard free energies of formation $\Delta_f G$ were taken from the 
literature \cite{alberty_biochemical_2006}. For other internal metabolites, for which such data does not exist, we employed a variant of the group contribution 
method \cite{Mavrovouniotis91, Mavrovouniotis90, Jankowski2008}. For each such molecule $g_1g_2\dots g_n$, composed from building 
blocks $\{g_i\}$, we calculated $\Delta_f G$ using
\begin{equation}
 \Delta_f G = E_0 + \sum_j E_1(g_j) + \sum_{<j,k>}E_2(g_j,g_k),
\end{equation}
where $E_0$ is a constant, $E_1(g_j)$ is the contribution of group $g_j$ and $E_2(g_j,g_k)$ is a small correction due to
neighbouring group-group interactions. The values of $E_0$, the vector $E_1$ and matrix $E_2$ are determined by performing a 
least-squares fit to a training set of molecules with known $\Delta_f G$s that correspond most closely to the linear CHOP 
molecules of our network (see Supporting Information \cite{suppmatt} section III). Thermodynamic profiles in Figs \ref{fig:TrunkPathway} and \ref{fig:Glyc_noConcRest} are plotted 
for 1M concentrations of all metabolites.

\subsection{Flux calculation}
We used the method of Ref.~\cite{Heinrich97} to calculate the flux carried by a linear pathway. This method assumes that the flux through 
reaction $i$ is given by \cite{Albery76, Heinrich97}:
\begin{equation}
 v_i = \frac{k_d [E_i] ( [S_{i-1}] q_i - [S_i] )}{ 1+q_i },
\end{equation}
where $k_d$ is the diffusion-controlled rate constant, $[E_i]$ is the enzyme concentration, $[S_{i-1}]$ and $[S_i]$
represent substrate and product concentrations and $q_i$ is the thermodynamic constant. This expression assumes that the enzyme acts as a perfect 
catalyst, and is used to derive an expression for pathway flux and metabolite concentrations
(a complete explanation can be found in the Supporting Information \cite{suppmatt} section IV and V).
We then use Powell's method \cite{NumericalRecipes} to find the set of enzyme concentrations which maximize the flux subject to  the constraints
that (1) all steady-state intermediate concentrations are within the prescribed range and (2) the total enzyme concentration is fixed.

\subsection{Sampling the parameter space}
We randomly selected 10000 points from the parameter space corresponding to the concentrations of 8 external metabolites and 
the G3P and pyruvate concentrations. Each parameter was sampled 
logarithmically over a range covering several orders of magnitude above and below its typical physiological concentrations (see Table~SIV).
For each of these parameter points, we calculated the optimized flux $J_i$ of 
each candidate pathway as detailed above, and computed the \emph{comparative flux} of path $i$ as $CF_{i} = J_i/\max{\{J_i\}}$, 
by dividing its flux  by the highest flux obtained across all pathways at the given point in parameter space.

\subsection{Robustness of our results to small free energy changes}
Using the group contribution method, the typical error in our calculation of the free energy of formation  $\Delta_f G$ for a given molecule is a 
few kJ/mol (see Supporting Information \cite{suppmatt} section III). 
To check the robustness of our results to such errors, our entire analysis was repeated using  $\Delta_f G$ values computed using different sets of  
training molecules, consisting of 80\% of 
the molecules from the original training set, chosen at random. 
The qualitative results using such networks were identical from those obtained from the full set of training compounds. 
For example, the top 25 glycolytic pathways 
obtained from the reduced set contained 23 out of the 25 pathways from the original analysis. Although the rank order of the pathways in terms of comparative 
flux across the whole parameter space did differ, the top 3 performing pathways were the same in both cases.

\section*{Acknowledgments}
BW and RJA contributed equally to this work. We thank Patrick Warren and Vincent Danos for many  helpful discussions. 
SJC was supported by a Carnegie/Caledonian Scholarship from 
the Carnegie Trust for Scotland, BW was supported by a Leverhulme Trust Early Career Fellowship and by a Royal Society of Edinburgh Personal Research Fellowship, 
and RJA was supported by a  Royal Society University Research Fellowship.

\section*{Conflict of interest}
The authors declare no competing financial interests.

\bibliography{refs}{}
\bibliographystyle{naturemag_no_url}

\end{document}